\documentclass[12pt]{article}
\usepackage[utf8]{inputenc}
\usepackage{amsmath,setspace,geometry}
\usepackage{amsthm}
\usepackage{amsfonts}
\usepackage[shortlabels]{enumitem}
\usepackage{rotating}
\usepackage{pdflscape}
\usepackage{graphicx}
\usepackage{bbm}
\usepackage[dvipsnames]{xcolor}
\usepackage{hyperref}
\hypersetup{colorlinks=true, linkcolor= BrickRed, citecolor = BrickRed, filecolor = BrickRed, urlcolor = BrickRed, hypertexnames = true}
\usepackage[]{natbib} 
\bibpunct[:]{(}{)}{,}{a}{}{,}
\usepackage[english]{babel}
\usepackage{float}
\usepackage{caption}
\usepackage{subcaption}
\usepackage{booktabs}
\usepackage{pdfpages}
\usepackage{threeparttable}
\usepackage{lscape}
\usepackage{bm}

\renewcommand*{\epsilon}{\varepsilon}

\usepackage{setspace}
\title{An MPEC Estimator for the Sequential Search Model}
\author{Shinji Koiso\thanks{koiso-shinji970@g.ecc.u-tokyo.ac.jp, Department of Economics, University of Tokyo}, Suguru Otani\thanks{suguru.otani@e.u-tokyo.ac.jp, Market Design Center, Department of Economics, University of Tokyo\\Declarations of interest: none}}
\date{
\today}

\begin{document}

\maketitle

\begin{abstract}
\noindent
This paper proposes a constrained maximum likelihood estimator for sequential search models, using the MPEC (Mathematical Programming with Equilibrium Constraints) approach. This method enhances numerical accuracy while avoiding ad hoc components and errors related to equilibrium conditions. Monte Carlo simulations show that the estimator performs better in small samples, with lower bias and root-mean-squared error, though less effectively in large samples. Despite these mixed results, the MPEC approach remains valuable for identifying candidate parameters comparable to the benchmark, without relying on ad hoc look-up tables, as it generates the table through solved equilibrium constraints.
\end{abstract}

\section{Introduction}

The consumer search process, through which individuals gather information about choices, is essential to understanding decision-making behavior. This process has become increasingly observable to researchers through browsing data, which reveals the options considered before a final choice. The availability of such data has allowed for the estimation of structural models of consumer behavior, as noted by \cite{ursu2023sequential}. The current benchmark model solves implicit functions related to reservation prices using a fixed-point approach, which is computationally demanding.

To address this, we propose an estimator based on the Mathematical Program with Equilibrium Constraints (MPEC) approach \citep{su2012constrained}. MPEC is a constrained optimization problem subject to equilibrium conditions, that avoids iterative solutions to the fixed-point problem and removes approximation and estimation errors — a key issue in demand estimation \citep{dube2012improving}, dynamic programming \citep{su2012constrained}, and misclassification models \citep{lu2014mpec}. 

Monte Carlo simulations show that MPEC performs comparably, with lower bias and root-mean-squared error (RMSE) in small samples but higher bias and RMSE in larger samples, relative to the common estimation method using an ad hoc look-up table. The MPEC approach is particularly valuable for identifying parameters comparable to the benchmark, eliminating the need for the look-up table. MPEC can
effectively construct this table by requiring the equilibrium constraints to be satisfied during estimation. We conclude that MPEC is a useful alternative for obtaining benchmark estimates before advancing to more complex models, especially when approximation and estimation errors from an ad hoc look-up table are unknown to researchers.

\section{Weitzman's sequential search model}

\subsection{Framework}
We construct the sequential search model based on \cite{weitzman1979optimal}.
A decision maker $i$ faces a set of boxes $\mathcal{J}=\{1,\cdots,J\}$ and box $j$ gives a potential reward $u_{ij}\in \mathbb{R}$ independently drawn from a known
distribution $F_{ij}(u)$.
Opening box $j$ takes cost $c_{ij}\in \mathbb{R}^{++}$. 
An outside option is denoted as $j = 0$ with a known reward $u_{i0}$ is available at no cost.
The decision maker
opens boxes via sequential search steps and her goal is to maximize her expected reward net of total costs.

Suppose that the decision maker has opened a set \(S_{i}\) of boxes,
which revealed a maximum reward value of \(u_{i}^{*}=\max _{j \in S_{i} \cup 0} u_{i j}\), and \(\bar{S}_{i}\) unopened boxes can still be
opened. 
Her dynamic programming problem choosing whether to stop opening boxes and get payoff \(u_{i}^{*}\), or to continue opening boxes is described by the following Bellman equation:
\begin{align}
    V\left(\bar{S}_{i}, u_{i}^{*}\right)=\max \left\{u_{i}^{*}, \max _{j \in \bar{S}_{i}}\left\{-c_{i j}+W_{j}\left(\bar{S}_{i}, u_{i}^{*}\right)\right\}\right\}\nonumber
\end{align}
where \(W_{j}\left(\bar{S}_{i}, u_{i}^{*}\right)\) is the expected value of continuing to open boxes and is
defined as
\begin{align}
    W_{j}\left(\bar{S}_{i}, u_{i}^{*}\right)=V\left(\bar{S}_{i} \backslash j, u_{i}^{*}\right) \int_{-\infty}^{u_{i}^{*}} d F_{i j}(u)+\int_{u_{i}^{*}}^{\infty} V\left(\bar{S}_{i} \backslash j, u\right) d F_{i j}(u).\nonumber
\end{align}
The reservation utility of a product \(z_{i j}\) is the utility level defined as 
\begin{align}
    \int_{z_{i j}}^{\infty}\left(u_{i j}-z_{i j}\right) d F_{i j}\left(u_{i j}\right)=c_{i j}.\nonumber
\end{align}

A set of optimal decision rules, developed by
\cite{weitzman1979optimal}, is used to characterize consumers' optimal search and choice strategies. 
The rules are as follows:
\begin{enumerate}
    \item Consumers know the true distribution(s) \(F_{i j}(u)\).
    \item Search fully reveals the utility associated with product \(j\).
    \item For each consumer \(i, u_{i j}\) is independently (across \(j\) ) drawn from \(F_{i j}(u)\).
\end{enumerate}
Then, the optimal search and choice decision
rules are expressed as follows:
\begin{enumerate}
    \item Selection Rule: The consumer searches in decreasing order of reservation utilities.
    \item Stopping Rule: Search terminates when the maximum observed utility exceeds the reservation utility of any unsearched product.
    \item Choice Rule: Once the consumer stops searching, she chooses the product with the highest observed utility among all searched options.
\end{enumerate}

\subsection{Parametrizations}

Empirical economists often assume consumer $i$'s utility defined as
\begin{align}
    u_{i j} & =\delta_{i j}+\varepsilon_{i j}  =\left(\xi_{i j}+\mu_{i j}\right)+\varepsilon_{i j},\nonumber\\
    \quad \quad \varepsilon_{i j}&\sim_{i.i.d} N(0,\sigma_{\mu}),\quad \mu_{i j} \sim_{i.i.d} N(0,\sigma_{\varepsilon})\nonumber
\end{align}
where \(\delta_{i j}\) is utility which is known by the consumer prior to search (``pre-search
utility") and \(\varepsilon_{i j}\) is utility that is only known by the consumer
after search (``post-search taste shock"). 
We assume that the pre-search utility
\(\delta_{i j}\) consists of a component \(\xi_{i j}\) that can be observed by the researcher and a pre-search taste shock \(\mu_{i j}\) that
cannot be observed by the researcher. 
According to \cite{ursu2023sequential}, we need to further normalize their variance by setting $\sigma_{\mu}= \sigma_{\varepsilon} =1$. 

Under the assumption of normally distributed post-search taste shocks, we can derive the following
expression for the reservation utility:
\begin{align}
    z_{i j}=\delta_{i j}+m\left(c_{i j}\right)=\xi_{i j}+\mu_{i j}+m\left(c_{i j}\right) \nonumber
\end{align}
where \(m\left(c_{i j} \right)\) is the implicit function that solves the following equation (see \cite{kim2010online}):
\begin{align}
    c_{i j}=\phi(m)+m \times[\Phi(m)-1] \label{eq:equilibrium_constraint}
\end{align}
with \(\phi\) and \(\Phi\) denoting the standard normal pdf and cdf, respectively. \cite{weitzman1979optimal} shows the existence and uniqueness of the solution of \eqref{eq:equilibrium_constraint}.

There are four primary methods to solve \eqref{eq:equilibrium_constraint}. The first method, proposed by \cite{kim2010online}, involves pre-computing the mapping between \( m \) and \( c \) and storing it in a look-up table. The second method, suggested by \cite{jiang2021consumer}, employs Newton's method to compute reservation utilities by iteratively improving approximations to the root of the function:

\[
    q(m) = (1-\Phi(m))\left(\frac{\phi(m)}{1-\Phi(m)} - m\right) - c = 0.
\]

The third approach, introduced by \cite{elberg2019dynamic}, uses a contraction mapping defined as:

\[
    \Gamma(m) = -c + \phi(m) + m \times \Phi(m).
\]

The fourth method, proposed by \cite{morozov2023measuring}, directly estimates \( m(c_{ij}) \).

\cite{ursu2023sequential} highlight limitations in each method: (1) the first method introduces errors due to linear interpolation for search cost values that do not align with grid points; (2) the second and third methods avoid interpolation errors but require iterative computation of \( m \) and a convergence threshold, which can cause numerical errors if the threshold is too loose; and (3) the fourth method involves estimation errors for \( m(c_{ij}) \). In practice, the second and third methods typically converge quickly, allowing for tight convergence thresholds that minimize numerical issues \citep{ursu2023sequential}. Similar challenges in demand estimation are addressed by the MPEC approach.

\section{An MPEC estimator for the sequential search model}
As a fifth method, we propose a straightforward estimator for the sequential search model, utilizing the Mathematical Programming with Equilibrium Constraints (MPEC) approach introduced by \cite{su2012constrained}. 
The MPEC estimator bypasses the need for iterative computations to find the fixed point by treating the equilibrium equations as constraints.

Let \(\theta\) represent the set of parameters. The MPEC estimator solves the following constrained optimization problem:
\begin{align}
    \max_{\theta}& \sum_{i\in \mathcal{N}} \log L_{i}(\theta,(z_{ij})_{j\in \mathcal{J}},(u_{ij})_{j\in \mathcal{J}})\nonumber\\
    \text{s.t.}\quad u_{i j} & =\xi_{i j}+\mu_{i j}+\varepsilon_{i j}\label{eq:mpec_formula}\\
    z_{i j}&=\xi_{i j}+\mu_{i j}+m\left(c_{i j}\right) \nonumber\\
    c_{i j}&=\phi(m)+m \times[\Phi(m)-1] \nonumber
\end{align}
where individual likelihood $L_{i}(\theta,(z_{ij})_{j\in \mathcal{J}},(u_{ij})_{j\in \mathcal{J}})$ is derived as
\begin{align}
    L_{i}(\theta,(z_{ij})_{j\in \mathcal{J}},(u_{ij})_{j\in \mathcal{J}}) &= \Pr (\underbrace{z_{ih} \geq \max_{k \in \mathcal{J}\setminus \{1,\cdots,h\}} z_{ik} \: \forall h \in S_i}_{\text{selection rule}}\nonumber\\
    & \cap \underbrace{z_{ih} \geq \max_{k = 1}^{h-1} u_{ik} \: \forall h \in S_i \cap \max_{h \in S_i \cup \{0\}} u_{ih} \geq \max_{l \in \bar{S}_i} z_{il} }_{\text{stopping rule}}\nonumber\\
    & \cap \underbrace{u_{iy_i} \geq \max_{h \in S_i \cup \{0\}} u_{ih} }_{\text{choice rule}}). \label{eq:likelihood}
\end{align}

A remarkable advantage of MPEC is that it does not need an ad hoc look-up table which is unknown to researchers and does not incorporate approximation and estimation error of equilibrium constraints \eqref{eq:equilibrium_constraint}, in addition to the main advantage of MPEC that it does not need to solve the fixed point problem iteratively. Detailed construction is provided in the Appendix.

\section{Simulation}

For comparison, we follow the parameter settings for \(\xi_{ij}\) and \(c_{ij}\) in \eqref{eq:mpec_formula} as described in Appendix B of \cite{ursu2023sequential}. We evaluate the MPEC approach against their kernel-smoothed frequency estimator (benchmark), which employs a look-up table—commonly used in empirical research. The same ad hoc table from \cite{ursu2023sequential}, with a grid fineness of 0.001, is utilized.
We generate 50 simulated datasets, each representing \(N\in\{500,1000\}\) consumers who make sequential search and purchase decisions across four brands and an outside option (with the mean utility of the outside option normalized to zero). The utility function includes only brand intercepts, specified as \((\beta_1,\beta_2,\beta_3,\beta_4) = (1.0, 0.7, 0.5, 0.3)\). The search cost logarithm is set at \(\log c = -3.0\), and \(D = 100\) draws are used for the error terms. All estimations start from an initial vector of zeros. The replication code, written in Julia for fair comparison, is available on the authors' GitHub.

\begin{table}[!htbp]
  \begin{center}
      \caption{MPEC vs ad hoc look-up (benchmark)}
      \label{tb:results_MPEC_1_bias_rmse} 
      \subfloat[$N=500$]{
\begin{tabular}[t]{llrrr}
\toprule
\multicolumn{1}{c}{ } & \multicolumn{2}{c}{MPEC} & \multicolumn{2}{c}{Look-up table} \\
\cmidrule(l{3pt}r{3pt}){2-3} \cmidrule(l{3pt}r{3pt}){4-5}
  & Bias & RMSE & Bias & RMSE\\
\midrule
$\beta_{1}$ & -0.179 & 0.250 & -0.212 & 0.273\\
$\beta_{2}$ & -0.156 & 0.216 & -0.168 & 0.237\\
$\beta_{3}$ & -0.044 & 0.071 & -0.109 & 0.200\\
$\beta_{4}$ & -0.062 & 0.182 & -0.060 & 0.192\\
$\log c$ & 0.214 & 0.250 & 0.248 & 0.304\\
Time &  & 67.957 &  & 12.691\\
\bottomrule
\end{tabular}
}
      \subfloat[$N=1000$]{
\begin{tabular}[t]{llrrr}
\toprule
\multicolumn{1}{c}{ } & \multicolumn{2}{c}{MPEC} & \multicolumn{2}{c}{Look-up table} \\
\cmidrule(l{3pt}r{3pt}){2-3} \cmidrule(l{3pt}r{3pt}){4-5}
  & Bias & RMSE & Bias & RMSE\\
\midrule
$\beta_{1}$ & -0.272 & 0.361 & -0.203 & 0.251\\
$\beta_{2}$ & -0.199 & 0.332 & -0.114 & 0.174\\
$\beta_{3}$ & -0.173 & 0.255 & -0.088 & 0.157\\
$\beta_{4}$ & -0.112 & 0.258 & -0.036 & 0.159\\
$\log c$ & 0.161 & 0.255 & 0.265 & 0.290\\
Time &  & 182.293 &  & 40.294\\
\bottomrule
\end{tabular}
}
  \end{center}
  \footnotesize
  Note: The benchmark results closely replicate Column (4) in Table B1: Monte Carlo Simulation Results from \cite{ursu2023sequential}. We calculate the average finish time for locally solved cases.
\end{table} 

Table \ref{tb:results_MPEC_1_bias_rmse} presents the bias and RMSE of the estimated coefficients. Panel (a) shows that the MPEC estimator, though still biased, achieves a smaller bias and RMSE than the benchmark in small samples, aligning with the MPEC misclassification model \citep{lu2014mpec}. However, Panel (b) reveals worse performance in both bias and RMSE for larger samples, except for the search cost. MPEC also requires over four times the computational time and struggles with finding local optima.

Despite these seemingly discouraging results, we argue that the MPEC approach remains useful for identifying candidate parameters comparable to those from the benchmark method, which relies on an ad hoc look-up table. Furthermore, MPEC can construct this table dynamically by solving the equilibrium constraints during the estimation process.

\section{Conclusion}

The optimal sequential search model, based on \cite{weitzman1979optimal}, has been widely used in empirical research \citep{ursu2023sequential}. However, caution is needed regarding estimation and approximation accuracy, as the commonly used approach relies on an ad hoc look-up table.

To address these issues, we propose an MPEC estimator that bypasses the need for approximations and estimation of equilibrium constraints. Despite certain limitations, the MPEC approach proves useful for identifying parameters comparable to the benchmark while dynamically generating the look-up table during the estimation process.

\paragraph{Acknowledgments}
This work was supported by JST ERATO Grant Number JPMJER2301, Japan.

\bibliographystyle{aer}
\bibliography{sequential_search}

\newpage
\appendix
\section{Appendix (for online publication)}

\subsection{Crude estimator}

We first introduce a crude estimator for likelihood expression \eqref{eq:likelihood} as the simplest approach.
Define
\begin{align}
    v_{i,1h} &= z_{ih} - \max_{k \in \mathcal{J}\setminus \{1,\cdots,h\}} z_{ik} \label{v1}\\
    v_{i,2h} &=  z_{ih} - \max_{k = 1}^{h-1} u_{ik} \label{v2} \\
    v_{i,3} &= \max_{h \in S_i \cup \{0\}} u_{ih} - \max_{l \in \bar{S}_i} z_{il} \label{v3}\\
    v_{i,4} &= u_{iy_i} - \max_{h \in S_i \cup \{0\}} u_{ih}  \label{v4}
\end{align}
Then, the estimation procedure is described as follows.
\begin{itemize}
    \item[1] Take $d = (1, \cdots, D)$ sets of draws of $\mu_{ij}$ and $\epsilon_{ij}$ (each set of draws contains one draw of $\mu_{ij}$ and one draw of $\epsilon_{ij}$) for each consumer-product combination, i.e., $D\times J \times N$ sets of draws.
    \item[2] For a given guess of parameters $\theta$, compute $u^d_{ij}$ and $z^d_{ij}$ for each set of draws $d$ and each consumer-product combination.
    \item[3] Calculate the expressions in equations \eqref{v1} to \eqref{v4} for each set of draws $d$ and each consumer. Compute the likelihood contribution for each consumer and draw:
    \begin{align*}
        L_i^d = \left[\prod_{h \in S_i} \bm{1}\{v^d_{i,1h} \geq 0\} \right] \times \left[\prod_{h \in S_i} \bm{1}\{v^d_{i,2h} \geq 0\} \right] \times  \bm{1}\{v^d_{i,3} \geq 0\}  \times  \bm{1}\{v^d_{i,4} \geq 0\}
    \end{align*}
    \item[4] Compute $L_i = \frac{1}{D}\sum_{d = 1}^D L_i^d$ for each consumer.
    \item[5] Compute $\log L = \sum_{i = 1}^N \log(L_i)$ and solve the constrained problem \eqref{eq:mpec_formula}.
\end{itemize}

\subsection{Kernel estimator}
To improve upon the crude estimator, the kernel estimator applies a smooth kernel function to obtain the log-likelihood. Specifically, we use a multivariate scaled logistic cumulative distribution function as the kernel, resulting in the following consumer-specific likelihood contribution:
\begin{align*}
    L_i^d = \frac{1}{1 + \sum_{k=1}^2 \sum_{h \in S_i} \exp (- \rho_k v^d_{i,kh}) + \sum_{k=1}^2 \exp (- \rho_k v^d_{i,k})},
\end{align*}
where \(\rho_k\) is a scaling parameter for each condition, to be determined by the researcher. The procedure to estimate \(v_k\) is the same as in the crude estimator. In our simulation, \(\rho_k\) is set to 10 for \(N=500\) and 20 for \(N=1000\) for both approaches. Further methods requiring fine-tuning are discussed in \cite{ursu2023sequential}.



\end{document}